\documentstyle[12pt,epsfig]{article}
\date{}
\title{On matrix elements of phase-angular momentum
commutator
in Hilbert space of arbitrary dimensions}
\author{{Ramandeep S. Johal}\\
{\it Institut f\"{u}r Theoretische Physik,}\\
{\it Technische Universit\"{a}t Dresden, }\\
{\it  01062 Dresden, Germany.}\\
e-mail: rjohal$@$theory.phy.tu-dresden.de\\
Ph.:  + 49 (0351) 463 35582  \\
Fax:  + 49 (0351) 463 37299}
\begin{document}
\baselineskip 24pt
\maketitle
\def\be{\begin{equation}}
\def\ee{\end{equation}}
\def\ba{\begin{eqnarray}}
\def\ea{\end{eqnarray}}
\begin{abstract}
We discuss correspondence between  the predictions of
quantum theories for rotation angle formulated in
infinite and finite dimensional Hilbert spaces, taking
as example, the calculation of matrix elements of
phase-angular momentum commutator. 
A new derivation of the matrix elements  is presented in infinite space,
 making  use of a unitary transformation that maps from
the state space of periodic functions to non-periodic
functions, over which the spectrum of angular momentum operator
is in general, fractional. The approach can be applied to
finite dimensional Hilbert space also, for which identical
matrix elements are obtained.
\end{abstract}

%PACS
%Keywords:
\newpage
Many interesting quantum phenomena, such as  
 in Josephson junctions \cite{jj}, Aharonov-Bohm effect 
\cite{ab}, fractional statistics in quantum hall effect \cite{aro} 
 or superfluid thin films \cite{lymr},
 are a consequence of the periodic or non-periodic nature of 
wave function under  change in the phase variable associated
with it. However,     
the idea of a hermitian phase operator, which may serve as possible  
observable in quantum mechanics, has a long history.
Dirac \cite{dirac} first of all attempted to obtain a phase
operator for harmonic oscillator by defining a
polar decomposition of annihilation operator,
$a = e^{i\hat{\phi}_D}\sqrt{a^{\dag}a}$. 
However, the exponential phase operator so defined
is not unitary \cite{sg} and so does not yield a hermitian
operator $\hat{\phi}_D$.
It is now generally accepted that a  hermitian
phase operator does not exist for harmonic oscillator. 
A related system is of a quantum rotor in plane.
Here due to the angular momentum being unbounded,
a hermitian operator $\hat{\phi}$ may be defined,
which has continuum of phase eigenvalues $\phi$. The Hilbert space of
the system is space of square integrable 
functions of polar angle $\phi$ in the range
$[0,2\pi]$, with Lebesgue measure $d\phi$. 
A differential realization for the third component
of angular momentum $\hat{L}_z =-i{\partial
\over \partial \phi}$, then implies the canonical
commutator $[\hat{\phi},\hat{L}_z] =i$. 
 
However to deal with bounded operators \cite{santh} 
or finite spin systems \cite{gold}, 
finite dimensional  version of phase operator
formalism have been proposed. A distinct feature 
of this formalism is that eigenvalues of phase operator
are restricted to quantised values.
Naturally then, differential realization
for $\hat{L}_z$ is not valid as it assumes a 
continuity of phase eigenvalues. A hermitian phase operator 
$\hat{\phi}_{l}$
so defined, satisfies a different commutator with $\hat{L}_z$ 
\be
[\hat{\phi}_{l},\hat{L}_z] = \frac{2\pi}{2l+1}
\sum_{m, m^{\prime} =-l
}^{l}\frac{(m^{\prime}-m)|m^{\prime}\rangle\langle m|}
{{\rm exp}[2\pi i(m-m^{\prime})/(2l+1)]-1},\quad (m\ne m^{\prime})
\label{com}
\ee
where $(2l+1)$ is dimension of the space and Dirac notation
for states is used. 
One such formalism that has gained attention in recent years, is
the Pegg-Barnett (PB) theory \cite{bp}. It emphasizes
a specific limiting procedure: the
physical results are to be obtained when the infinite dimensional
limit is taken, {\it after} the calculation of expectation values 
for observables. For the angular momentum case, it means that only in the
semi-classical limit ($l\to \infty$), can predictions of the theory
match with physical results.  Although the analogous
formalism for phase of quantised electromagnetic field has
been intensively studied and also debated in recent years \cite{lynch},
there have been relatively fewer studies elaborating the
implications of the PB theory for rotation angle in quantum
mechanics.  An example is non-trivial notion
of angular velocity in this formalism, where the usual
Heisenberg equation for phase operator does not predict  
physically intuitive result. A solution was presented in \cite{rj99}
 by recourse to the classical-quantum correspondence. 

The infinite space formalism has its own share of subtleties.
For example, the usual Heisenberg uncertainty relation between 
$\hat{\phi}$ and $\hat{L}_z$ does not apply \cite{kraus}. 
One of the often discussed issues  
has been the seeming inconsistency in the calculation of 
matrix elements of canonical commutator,   
$(n,[\hat{\phi},\hat{L}_z]m)$, over  
 periodic wave functions $u(\phi)\propto e^{im\phi}$, satisfying   
$u(\phi +2\pi) =u(\phi)$. 
It was thought that the problem lies in
the operation $\hat{\phi}u(\phi) =\phi u(\phi)$,
which projects wave functions out of the space
of periodic functions. So the use
of a bare phase operator was rejected in favour of 
defining a suitable periodic extension of phase operator \cite{sg,jule}
 or using periodic functions of the  phase operator \cite{cn68}.
However, if the adjoint of $\hat{L}_z$ is taken carefully, then 
the problem vanishes \cite{lm} and  well
behaved matrix elements of the commutator are obtained
\be
(n,[\hat{\phi},\hat{L}_z]m)={1\over 2\pi}\int_{0}^{2\pi}d\phi\;
e^{-in\phi}[\hat{\phi},\hat{L}_z]
e^{im\phi} =i\delta_{mn}.
\label{rlim}
\ee
However, within the finite dimensional space, different
matrix elements are obtained \cite{bp}. For instance, given the commutator
(\ref{com}), the diagonal elements vanish
\be
\langle m|[\hat{\phi}_{l},\hat{L}_z]|m\rangle=0.
\label{diag}
\ee
This result is exact for all angular momentum eigenstates $|m\rangle$ 
where  $-l\le m\le l$. Thus 
taking $l\to \infty$, does not yield the result in
infinite space. In this paper, we try to address the  
question of correspondence between results of these matrix 
elements in the finite and infinite dimensional spaces.
We adopt a different approach for the calculation of these
elements and argue that results in finite space can also
match with those in the infinite space.

The difficulty discussed above for infinite space springs 
from the fact that $\hat{L}_z$
is not self-adjoint on the space of nonperiodic functions
of the kind $\phi u(\phi)$ which turn up during the calculation.
Still, this does not exclude the use of other 
non-periodic functions (if they serve any useful purpose)
 over which angular momentum operator is self-adjoint. 
Note that using the differential realisation
and by partial integration, we obtain
\be
(u_2,\hat{L}_zu_1) 
 = (\hat{L}_zu_2,u_1) -{i\over 2\pi}
 [{u_2}^*(2\pi)u_1(2\pi) - {u_2}^*(0)u_1(0)].
\ee
The surface term vanishes or alternately,
 $\hat{L}_z$ becomes hermitian  if its eigenfunctions
satisfy $u_s(2\pi) = e^{i2\pi s}u_s(0)$, where   
$s$ is positive real parameter. 
We denote the angular momentum operator by $\hat{L}_z(s)$,
 whose domain ($D$) is the
above mentioned space of non-periodic functions $u_s(\phi)$.
It is clear that $D[\hat{L}_z]\ne D[\hat{L}_z(s)]$ for 
non-integer values of parameter $s$,
and that both operators have same differential realisation 
$-i\partial /{\partial \phi}$.  
In other words, both satisfy the same canonical commutator
with $\hat{\phi}$ and the matrix elements of this commutator
can be correctly evaluated over their respective domains along the lines
presented in \cite{lm}.
In literature, usually hermiticity is guaranteed for $\hat{L}_z$
by restricting its state space to periodic functions ($s$ is zero or integer
valued). But it is clear that periodicity of wave functions is
required only to keep the spectrum of $\hat{L}_z$ integer valued.
Theoretical  possibilities such as generalized statistics do exist,
 where spectrum of $\hat{L}_z$ is in general, fractional \cite{aro,lymr}.
In the following, we make  use of the freedom of parameter $s$ to 
derive the matrix elements of canonical commutator.
 
We have seen that $D[\hat{L}_z(s)]$ consists of wave functions   
$u_s(\phi) \propto e^{i(m+s)\phi}$. These can be generated from
the usual periodic functions $u(\phi)$, by a unitary transformation:
$e^{is\hat{\phi}}u(\phi) = u_s(\phi)$.  
Also we have
$\hat{L}_z(s)u_s(\phi) = (m+s)u_s(\phi)$,
in analogy with $\hat{L}_zu(\phi) =m u(\phi)$.
Now consider the term 
\be
\hat{R} = e^{is\hat{\phi}}\hat{L}_z -\hat{L}_z(s)e^{is\hat{\phi}}.
\label{rterm}
\ee
Evaluating $\lim_{s\to 0} {1\over is} (n,\hat{R}m)$, 
we can easily verify that (\ref{rlim}) is satisfied.
Note that the need for taking adjoint of 
angular momentum operator is dispensed with,  
since $\hat{L}_z$ and  $\hat{L}_z(s)$ have to act 
in their respective domains only.

It is the unitary transformation $e^{is\hat{\phi}}$ which is of present 
interest to us. For instance, it can be used to define 
a generalized dynamics \cite{rj00} for current-biased Josephson junction.
The parameter $s$ in this case, takes a physical interpretation
and represents an external magnetic flux linked to the junction.
This system is equivalent to the system of a charged particle
on a ring with magnetic flux through the centre of the ring.
Thus the limit $s\to 0$ in our calculation, represents the  limit
in which magnetic flux at the centre of ring goes to zero.

 For transparency, we also calculate in Dirac notation.
 The states
$\{|m+s\rangle \}$ form a complete orthonormal basis like the 
standard angular momentum basis $\{|m\rangle \}$.
The periodic phase eigenstate  can be written as
\be
|\phi\rangle = {1\over \sqrt{2\pi}} \sum_{m=-\infty}^{+\infty}
e^{-im\phi}|m\rangle, \label{p1}
\ee
or  equivalently  
\be 
|\phi\rangle = {1\over \sqrt{2\pi}} \sum_{m=-\infty}^{+\infty}
e^{-i(m+s)\phi}|m+s\rangle. \label{p2}
\ee
Relevant operators are defined by the following 
representation
\ba
\hat{L}_z|m\rangle &=& m|m\rangle,\\
\hat{L}_z(s)|m+s\rangle &=& (m+s)|m+s\rangle,\\
\hat{\phi}|\phi\rangle &=& {\phi}|\phi\rangle.
\ea
The representations (\ref{p1}) and (\ref{p2}) 
 imply that the following unitary transformation exists
\be
e^{is\hat{\phi}} = \sum_{m=-\infty}^{+\infty}
|m+s\rangle\langle m|. \label{uny}
\ee
The unitarity of this operator follows simply from
completeness of states.
For $s=0$, we get resolution of unity over 
standard angular momentum states.
Note again that due to the angular momentum
 being unbounded, the system of states is invariant
under any finite integer shift $s$. 
Due to this invariance, we can restrict the range of $s$
as $0\le s\le 1$.
Particularly, for $s=1$, we have the canonical  unitary
phase operator which shifts an angular momentum state
by unity. 
For this case, the unitary operator may be said to shift
states within the same orthonormal basis, while for
other values of $s$, (\ref{uny}) clearly takes
from one orthonormal basis to another.

Next we consider the matrix elements when dimension
of angular momentum space is finite.
From the $\hat{R}$ term (\ref{rterm})
defined in infinite space, we were able to derive the matrix elements
of canonical commutator $[\hat{\phi},\hat{L}_z]$. Thus it is interesting 
to extend this calculation to finite space also and see if we can recover,
 say (\ref{diag}) or not.
We take $\hat{R}= e^{is\hat{\phi}_l}\hat{L}_z-
\hat{L}_z(s)e^{is\hat{\phi}_l}$, and calculate
the matrix elements $\langle n|\hat{R}|k\rangle$, where
now all operators act in $(2l+1)$ dimensional space.
Analogous to (\ref{p1}) and (\ref{p2}),
we can define the phase states in finite space in terms of
either the  $\{|m\rangle\}_{-l,...,l}$ states
or $\{|m-1+s\rangle\}_{-l,...,l}$ states, as follows
\ba
|\phi_l\rangle &=& {1\over \sqrt{2l+1}} \sum_{m=-l}^{+l}
e^{-im\phi_l}|m\rangle, \nonumber\\
&=&
 {1\over \sqrt{2l+1}} \sum_{m=-l}^{+l}
e^{-i(m-1+s)\phi_l}|m-1+s\rangle. \label{pxy}
\ea
These are eigenstates of phase operator 
$\hat{\phi}_l|\phi_l\rangle = {\phi_l}|\phi_l\rangle$,
which form a conjugate $(2l+1)$-dimensional orthonormal basis,
if the phase eigenvalues are quantised as 
 $\phi_l = {2\pi n\over 2l+1}$, where $n =0,1,...,2l$, for phase operator
defined in $[0,2\pi)$ range.
Apart from $\hat{L}_z|m\rangle = m|m\rangle$,  we also have
\be
\hat{L}_z(s)|m-1+s\rangle = (m-1+s)|m-1+s\rangle.
\ee
From (\ref{pxy}), we have the following unitary transformation
\be
e^{is\hat{\phi}_l} = \sum_{m=-l}^{l-1}
|m+s\rangle\;\langle m| +|-l-1+s\rangle\langle l|,
\label{unyl}
\ee
which may be looked as a generalized phase operator
producing arbitrary shift in the angular momentum states.
Its operation on angular momentum states has been shown pictorially in
the Figure.  Note that for $s=1$, we have the standard phase operator 
formalism in finite space \cite{gold,bp}.  

Before we present results for matrix elements, a remark is 
in place. For infinite dimensional case, 
we can take the $s\to 0$ limit at the operator
level, for example,  in  (\ref{uny}). Thus one
recovers the canonical commutator 
$[\hat{\phi},\hat{L}_z]$ from  the quantity ${\hat{R}\over is}$, 
in the $s\to 0$ limit.  
However, this is not correct within finite space theory,
as can be seen from (\ref{unyl}); taking  $s\to 0$ limit 
does not give the identity operator. The reason 
for this difference is that 
in order to satisfy unitarity, the operator (\ref{unyl})
produces a shift of $s$ in all $|m\rangle$ states, except
for $m=l$ case, for which the shift is $\sigma = -2l-1+s$.
This discontinuity in the shift property of unitary
phase operator is also present in the $s=1$ case of standard
formalism \cite{bp}.
Note that  we should investigate the limit in which
magnitude of the shift produced goes to zero. In the infinite dimensional
case, shifts in all states were of equal magnitude. 
As an implication, separate limits are to be taken depending on
whether in the calculation of matrix elements, the state $|l\rangle$ 
is involved or not. Thus the said limit is
taken only {\it after} the matrix elements are calculated.

Now observing that
\be
\langle n|e^{is\hat{\phi}_l}\hat{L}_z|k\rangle =
k\sum_{m=-l}^{l-1}\langle n|m+s\rangle \delta_{mk}
+k \langle n|-l-1+s\rangle \delta_{lk},
\label{t1}
\ee
and 
\be
\langle n|\hat{L}_z(s)e^{is\hat{\phi}_l}|k\rangle =
\sum_{m=-l}^{l-1} (m+s)\langle n|m+s\rangle \delta_{mk} 
+(-l-1+s)\langle n|-l-1+s\rangle \delta_{lk}
\label{t2}
\ee
we can calculate $\langle n|\hat{R}|k\rangle$.
To evaluate proper limit, we distinguish two cases:

(i) If for some $m$ in the range $-l\le m \le l-1$,  $k=m$,
 then we obtain 
\be
\langle n|\hat{R}|k\rangle = -s\langle n|k+s\rangle.
\ee
Thus we get 
\be
\lim_{s\to 0} {1\over is} \langle n|\hat{R}|k\rangle = i\delta_{nk}.
\label{lim1}
\ee
(ii) If $k=l$, then  
\be
\langle n|\hat{R}|l\rangle = -\sigma \langle n|l+\sigma\rangle,
\ee
where parameter $\sigma$ has been defined above.
Therefore considering the limit when the shift $\sigma$
approaches zero 
\be
\lim_{\sigma \to 0} {1\over i\sigma} \langle n|\hat{R}|l\rangle 
= i\delta_{nl}.
\label{lim2}
\ee
Thus from our result for infinite dimensional space,
we might expect that limiting behaviour of 
$\langle n|\hat{R}|k\rangle$ in finite space, may yield 
the matrix elements such as (\ref{diag}).
 However, we see that the obtained matrix
elements (\ref{lim1}) and (\ref{lim2}) exactly correspond with
the elements  (\ref{rlim}),
provided the limits in which the shift parameter goes to  zero
are carefully taken.

Concluding, we have presented a new derivation for
matrix elements of canonical commutator between phase 
and angular momentum operators. We have utilised the more general 
state space of non-periodic eigenfunctions over which the spectrum of
angular momentum can be fractional. The matrix elements 
are obtained in the limiting case ($s\to 0$), when integer valued
angular momenta are realised. We remark that  
the limit $s\to 0$ has been used here only as
calculational tool; we do not imply that matrix elements
of the canonical commutator are well defined only for
$s\to 0$ case. In particular, we do not question
  the relevance  of canonical
commutator $[\hat{\phi},\hat{L}_z(s)] =i$ 
for circumstances where $s>0$ may be significant \cite{aro,lymr}. 
We have extended the technique to finite dimensional
space also, where it is required to introduce a generalization (\ref{unyl})
of unitary phase operator such that standard case \cite{bp} of $s=1$
is also recovered.  It is argued  that similar matrix elements as of
infinite space  can be
obtained. Thus we have here another example
of the correspondence between predictions  of 
 quantum  theories for rotation angle in infinite and finite
dimensional spaces.  
\section*{Acknowledgements}
The author is grateful to  Alexander von Humboldt Foundation,
Germany for financial support.
 
\newpage
\begin{figure}[htb]
\begin{center}
\epsfig{file=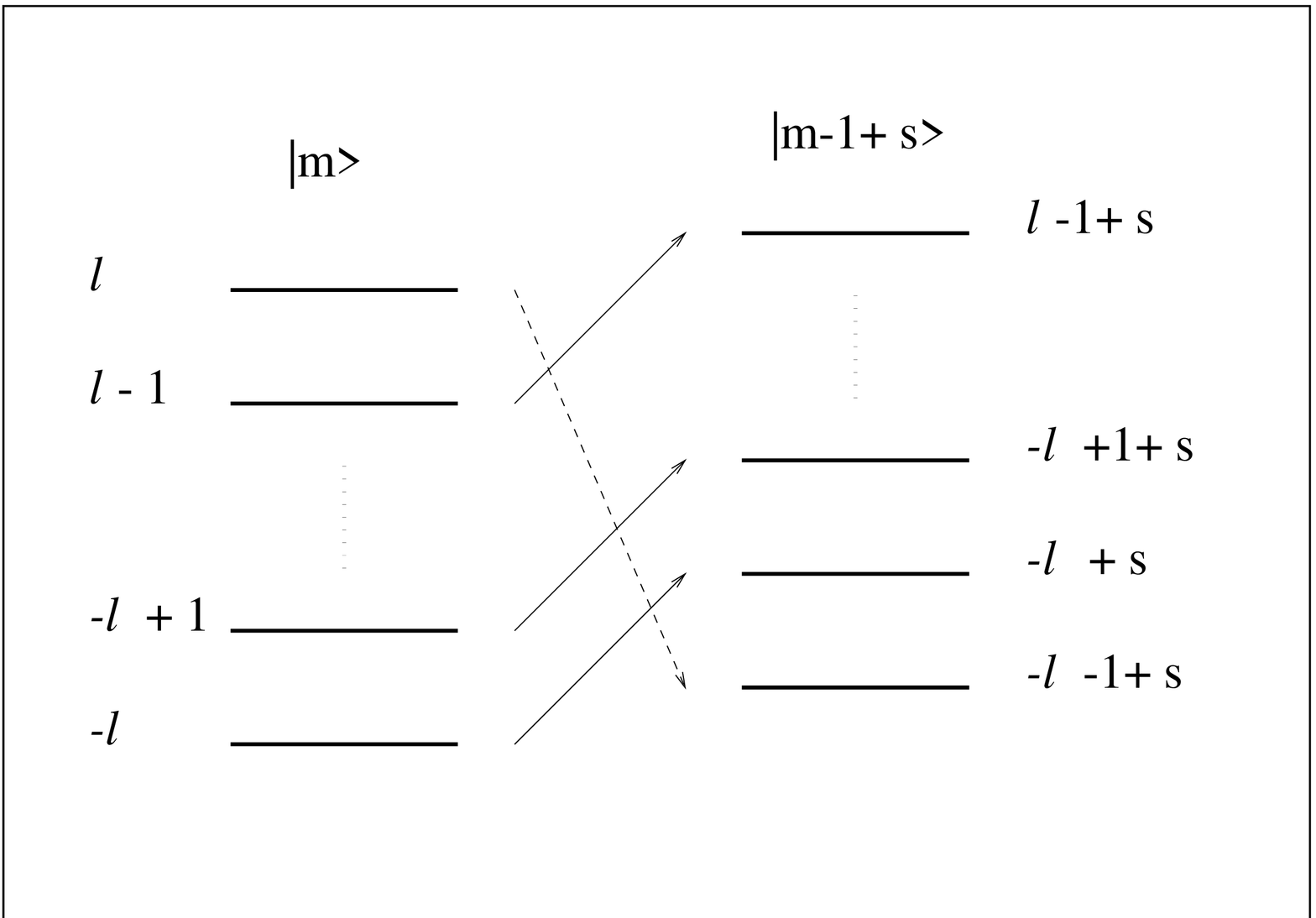,width=4in, angle =-90}
\caption{ The states $|m\rangle$ and $|m-1+s\rangle$
 of the $(2l+1)$-dimensional bases can be related by the 
 unitary transformation $e^{is\hat{\phi_l}}$
 of (\ref{unyl}), as shown by arrows. For $s=1$, 
the unitary operator of \cite{gold,bp} is obtained
which shifts the states by unity within basis $|m\rangle$.} 
\end{center}
\end{figure}
\end{document}